%% file: main.tex
\documentclass[sigconf,nonacm]{acmart}
\AtBeginDocument{%
  \providecommand\BibTeX{{%
    \normalfont B\kern-0.5em{\scshape i\kern-0.25em b}\kern-0.8em\TeX}}}

\usepackage{tikz}
\usepackage{amsmath}

\microtypecontext{spacing=nonfrench}

\usepackage[capitalize]{cleveref}
\usepackage{bm}
\usepackage{multirow}
\usepackage{algorithm} 
\usepackage{algpseudocode} 
\usepackage[caption=false]{subfig}
\usepackage{caption} 
\usepackage{comment}
\usepackage{balance}

\begin{document}

\title{Filtering DDoS Attacks from Unlabeled Network Traffic Data Using Online Deep Learning}

\author{%
  {\rm Wesley Joon-Wie Tann, Jackie Tan Jin Wei, Joanna Purba, \& Ee-Chien Chang} \\
  Department of Computer Science, National University of Singapore \\
  \texttt{$\{$wesleyjtann,jackietan,joanna\_sara$\}$@u.nus.edu,changec@comp.nus.edu.sg} \\
} %

\renewcommand{\shortauthors}{Tann et al.}

\input{section/abstract}




\maketitle

\input{section/sec1-intro}

\input{section/sec2-background}


\input{section/sec3-problem}


\input{section/sec4-associate}
\input{section/sec5-iterative}


\input{section/sec6-lstm_based}
\input{section/sec7-eval}

\input{section/sec8-related}


\input{section/sec9-conclude}



\bibliographystyle{ACM-Reference-Format}
\bibliography{bibliography}

\newpage
\appendix
\input{section/appendix.tex}



\end{document}

%% file: section/abstract.tex
\begin{abstract}

DDoS attacks are simple, effective, and still pose a significant threat even after more than two decades. Given the recent success in machine learning, it is interesting to investigate how we can leverage deep learning to filter out application layer attack requests. There are challenges in adopting deep learning solutions due to the ever-changing profiles, the lack of labeled data, and constraints in the online setting.
Offline unsupervised learning methods can sidestep these hurdles by learning an anomaly detector $N$ from the normal-day traffic ${\mathcal N}$. However, anomaly detection does not exploit information acquired during attacks, and their performance typically is not satisfactory. 
In this paper, we propose two frameworks that utilize both the historic ${\mathcal N}$ and the mixture ${\mathcal M}$ traffic obtained during attacks, consisting of unlabeled requests. We also introduce a machine learning optimization problem that aims to sift out the attacks using ${\mathcal N}$ and ${\mathcal M}$. 
First, our proposed approach, inspired by statistical methods, extends an unsupervised anomaly detector $N$ to solve the problem using estimated conditional probability distributions. We adopt transfer learning to apply $N$ on ${\mathcal N}$ and ${\mathcal M}$ separately and efficiently, combining the results to obtain an online learner.
Second, we formulate a specific loss function more suited for deep learning and use iterative training to solve it in the online setting. On publicly available datasets, our online learners achieve a 99.3\% improvement on false-positive rates compared to the baseline detection methods. In the offline setting, our approaches are competitive with classifiers trained on labeled data.

\end{abstract}

%% file: section/sec1-intro.tex
\section{Introduction}
Distributed Denial-of-Service (DDoS) attacks are well established as a significant threat to our present-day Internet network security, denying legitimate users access to shared and essential resources. The earliest reported DDoS attack in 1999~\cite{criscuolo2000distributed} started a wave of denial-of-service attacks that are distributed in nature. Even though more than two decades have passed, these attacks---simple to set up, difficult to stop, and very effective---are more popular than ever as they remain potent~\cite{ISTR16-04}.
The allocation of extra resources is a typical protective approach to handle such DDoS traffic. However, catering excess resources goes hand in hand with additional costs. There is a growing urgency to find practical and inexpensive methods, which can effectively abate disruptions by filtering malicious traffic during an attack. 
We focus on devising such a filtering system.

A primary form of DDoS defense, anomaly detection mechanisms~\cite{peng2007survey,liu2009botnet}, faces growing difficulty in detecting malicious traffic as application-layer attacks are increasingly sophisticated. There are several significant challenges. First, the attacks are often domain-specific with varying characteristics, making it hard for defense methods to generalize them. Hence a two-class classifier trained on historic attacks might not be applicable to different victims. Second, the time constraint is a consideration in the real-time environment. Online models have to be computationally efficient to respond to requests rapidly during an on-going attack. Third, while online machine learning seems to be an effective way to handle these evolving dynamics, the lack of labeled requests at the time of the attacks, required for supervised learning algorithms, poses another challenge for this learning paradigm. We must address these challenges before deploying them in the real world.

Existing DDoS defenses generally can be classified as statistical-based or machine learning-based detection. Statistical methods typically involve some fast calculation of network traffic properties, such as entropy scoring of network packets~\cite{7345297,6601602,8256833,8466805}, IP filtering~\cite{1204223,Peng02detectingdistributed}, and IP source traceback~\cite{10.1007/3-540-36159-6_4,5467062}. 
On the other hand, deep learning exploits intrinsic properties but could be computationally intensive. Most deep learning approaches for DDoS in the literature adopt offline learning, where the training is conducted much earlier (e.g., at least a few days) before the attack commences~\cite{8066291,doi:10.1002/dac.3497,Yuan2017DeepDefenseID,8666588,8416441}.
There are fewer works on online learning that performs incremental updates as more attack requests arrive. {Çakmakçı} et al. gave an up-to-date survey of online DDoS detection~\cite{DANESHGADEHCAKMAKCI2020102756}. Most online learning methods employed unsupervised clustering (e.g., k-means) on the attack day data~\cite{lima2019smart,7987186,DANESHGADEHCAKMAKCI2020102756}. 

In this paper, we explore how deep learning is leveraged to enhance detection performance, especially in the application-layer traffic, which exhibits intrinsic statistical properties. We formulate a two-class learning problem, which utilizes data from two operational periods, normal-day traffic $\mathcal{N}$, and the attack-day traffic $\mathcal{M}$ that contains a mixture of unlabeled legitimate and malicious traffic. We also assume a rough estimate $\alpha$ of the proportion of attack requests in $\mathcal{M}$. The learning objective is to accurately predict requests from $\mathcal{N}$ as normal, and $\alpha$ proportion of requests in $\mathcal{M}$ as attacks.

    \begin{figure}[tbp]
        \centering
        \includegraphics[width=.5\linewidth]{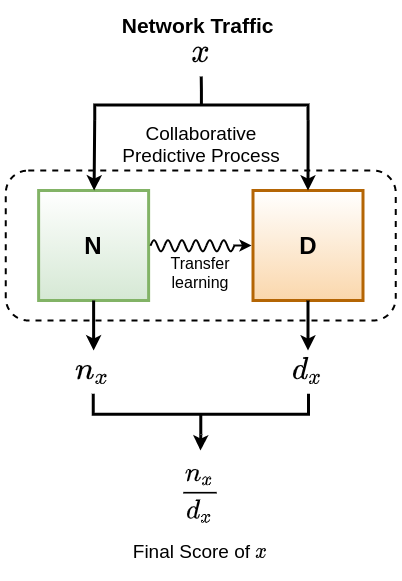}
        \caption{
        A framework that utilizes both ${\mathcal N}$ and ${\mathcal M}$ to rank network traffic based on their final $n_x / d_x$ scores. The higher the final score, the more likely the traffic is normal. To achieve computation speedup, the embedding of $N$ is transferred to $D$.
        }
        \label{fig:danframework}
    \end{figure}        

To solve this problem, we first employ techniques inspired by the statistical method PacketScore~\cite{1354679}, which takes a principled approach to estimate conditional probabilities (the likelihood that a request is an attack given that it is observed in the mixture traffic) of network requests. 
We perform the learning in a two-step process. (1) We apply unsupervised learning on the normal-day traffic $\mathcal{N}$ to obtain a model $N$ that learns the distribution $P_n$ of $\mathcal{N}$. (2) When an attack commences, a similar unsupervised learning model $D$ rapidly learns the distribution $P_d$ of the mixture $\mathcal{M}$. Now, the attack and normal requests can be differentiated based on the differences between two conditional probability distributions $P_n$ and $P_d$. We then rank each request $x$ by $P_n(x)/P_d(x)$, where a higher score indicates a higher chance of normality, as illustrated in Figure~\ref{fig:danframework}.

In our construction, both models $N$ and $D$ are LSTM recurrent neural networks, designed for application-layer DDoS, where a ``request'' consists of a sequence of events (e.g., an \texttt{HTTP GET} to the landing page, followed by subsequent \texttt{HTTP GET} requests). To meet the online requirements, we design model $D$ to achieve quick online updates by relying on transfer learning. When an attack occurs, the optimized embedding of $N$ is transferred to $D$, significantly reducing the training needed for $D$. We call this particular construction of online learning the \textit{N-over-D} nets. 

However, there are two main challenges of generalizing statistical methods to a deep-learning approach. First, the learning models normalize data values for effective training and smooth gradient flow, and even though the range of values is maintained, the resulting model-calculated probabilities are not exact. While we designed our LSTM to get close to the exact likelihood, it could be distorted when taking the $P_n/P_d$ division. Second, models $N$ and $D$ are separately trained during each operational period. There is no joint training of the models, and we do not fully leverage all available information for each model.

Hence, we formulate a machine learning loss function specifically to address these challenges. As N-over-D follows a statistical method to get probabilistic measures, it might not attain optimality in the learning paradigm. We design an online iterative two-class classifier for this particular loss function that jointly trains on both normal-day $\mathcal{N}$ and mixture $\mathcal{M}$ traffic.
This online classifier does not have the ground truth label information. It uses pseudo-ground-truth labels, taking all traffic during attacks as malicious and labeling them accordingly. 
We train the classifier, with a similar model architecture as the LSTMs in N-over-D, using these inexact labels to improve iteratively. This enhanced iterative classification approach achieves remarkable detection performance. 
Although the iterative classifier takes a slightly longer training time than N-over-D (3--5 times longer), it is nevertheless suitable for attacks with a short timeframe; even used for classifying unlabeled historic $\mathcal{M}$.


Comprehensive experiments demonstrate the ability of our approach to mitigate a range of DDoS attacks. For example, in the \textit{CICIDS2017} dataset, while the $N$ model achieves accuracy and a false-positive rate of (32.5\%, 69.97\%), both the online N-over-D and iterative classifier models accomplish marked improvements of (93.3\%, 0.47\%) and (93.5\%, 3.40\%), respectively. 
Moreover, both the N-over-D and the iterative classifier outperform existing online detection methods~\cite{lima2019smart,DANESHGADEHCAKMAKCI2020102756,Doriguzzi_Corin_2020} by a large margin in either accuracy or false-positive rates (see \cref{sec:experiments} for details). \\

\noindent\textbf{Contribution.} 
\begin{enumerate}
\item We formulate DDoS detection on unlabeled data as a machine learning optimization problem with normal-day ${\mathcal N}$ traffic, mixture ${\mathcal M}$ traffic, and the estimated $\alpha$ proportion of attack as training inputs. 

\item We propose an online approach N-over-D, inspired by statistical methods, extending one-class unsupervised learning to a two-class learning method. It deploys an LSTM-based model on ${\mathcal N}$ and ${\mathcal M}$ separately and incorporates transfer learning to speedup training in the online setting. 

\item We design a specific loss function to address the challenges of N-over-D, proposing an iterative online method for this loss function, which jointly trains on both normal-day $\mathcal{N}$ and mixture $\mathcal{M}$ traffic data.
\end{enumerate}

\noindent\textbf{Organization.} We organize the rest of the paper by first presenting the background information and challenges of the DDoS problem in \cref{sec:background}. We introduce the problem in \cref{sec:formulate} and propose an appropriate conditional probability learning approach in \cref{sec:pan}.
We then detail the iterative classification approach in \cref{sec:iterative} and the online LSTM architecture of our approaches in \cref{sec:associate}.
In \cref{sec:experiments}, we describe the datasets and evaluate our approaches on these datasets, comparing them with state-of-the-art methods. We also analyze the performance of our mitigation process and discuss several practical issues. Finally, we summarize related work in \cref{sec:relatedwork} and conclude the paper in \cref{sec:conclude}.


%% file: section/sec2-background.tex
\section{Background and Challenges}
\label{sec:background}
This section gives an overview of application layer network services 
and describes the challenges of detecting DDoS attacks in such systems. We also motivate the need for innovative defenses that can perform online network traffic monitoring.

\subsection{Application Layer DDoS}
A surge in application-layer DDoS attacks, using a network of widely distributed botnets, are increasingly disruptive to current network services~\cite{ISTR16-04}. Symantec's telemetry shows that it is often small and medium-sized retailers, selling goods ranging from clothing to gardening equipment to medical supplies, that are on the receiving end of the majority of attacks~\cite{ISTR19-02}. It is a serious global issue that could potentially affect any business that operates online, especially the essential services during critical times, such as the COVID-2019 pandemic. Observed in Q1 2020, there is an unexpected spike in the number of attacks; noticeable changes in the distribution of DDoS attacks by type doubled by $80\%$ against Q1 2019~\cite{KasperskyQ42019}.

\subsection{Existing Online DDoS Detection}
Several unsupervised learning mitigation strategies~\cite{lima2019smart,7987186,DANESHGADEHCAKMAKCI2020102756} have been proposed to counter such attacks in an online manner. 
These learning-based methods, employing rudimentary machine learning algorithms, e.g., random forest, logistic regression, decision tree (see Sec. \ref{subsec:ml} for details), perform clustering and simple distance measure using a few elementary features. While they can capture some basic network traffic properties, they are unable to learn deeper representations of legitimate and malicious requests to better distinguish between the two types of traffic. Moreover, the optimization goal of these methods is not clear.

\subsection{Flooding Attacks}
Common application-layer flooding attacks such as HTTP flood attacks are designed to generate an overwhelming amount of traffic, increasing the servers' load to severely impact a target. A popular tactic used by attackers to disrupt legitimate access is to disguise their traffic as flash crowds. A flash crowd is a surge in traffic to a particular Web site that causes the site to be virtually unreachable~\cite{6060809}. Even though attacks and flash crowds originate from different motivations, such attacks' covertness makes them hard to detect. These attacks typically send numerous HTTP packets abruptly and rapidly, flooding a target server with harmless HTTP requests, evading most defense systems. Any defense system designed to detect and mitigate such attacks must differentiate between attack and normal traffic fast, minimize losses, and perform online mitigation.

%% file: section/sec3-problem.tex
\section{Two-Class Learning on Unlabeled Mixture Traffic} 
\label{sec:formulate}   

In this section, we formalize the learning problem in the DDoS setting.
We consider two operational periods, a normal period  and  an attack period.  In practice, 
the attack period can be readily  detected as  it  is characterised by an unusual high volume of requests. 
Let $\mathcal{N}$ be the requests logged during the normal period,  and $\mathcal{M}$ the requests during the attack period.  The mixture $\mathcal{M}$ contains a  mixture of both legitimate  requests and attack requests.  We assume that based on the volume, we have a rough estimate of $\alpha$,  which is  the proportion of attack requests in  $\mathcal{M}$.   While $\mathcal{M}$ is unlabeled,  all requests in $\mathcal{N}$ are normal. 
Given $\mathcal{N}$, $\mathcal{M}$, and $\alpha$, the goal is to derive a two-class learning method that differentiates attack from normal, so that    $\mathcal{N}$ would be classified as  normal, and $\alpha$ proportion of $\mathcal{M}$  would be classified as attacks.
%



\paragraph{Offline vs Online Setting.}  
The online setting consists of  three stages. 
\begin{itemize}
\item In the {\em pre-processing}  stage,   computational  intensive processing can be performed on  $\mathcal{N}$, giving some intermediate model and representation $\widetilde{N}_0$. 
\item The {\em  on-line learning} stage is carried out when the attack commences. The time is divided into short intervals of $\ell$ minutes (e.g., $\ell=1$),  and
the mixture $\mathcal{M}$ is also divided according to $\mathcal{M}_1, \mathcal{M}_2, \ldots$, where 
 $\mathcal{M}_i$ contains  the mixture within the $i$-th interval. 
During the $i+1$-interval,  training is performed on    $\mathcal{M}_i$  and  some intermediate representation $\widetilde{N}_i$, giving the updated $\widetilde{N}_{i+1}$ and a model $C_{i+1}$.    There are stringent computing resource constraints, e.g., the training should be completed  within $\ell$ minutes.
\item
 In the {\em filtering} stage,  the trained model $C_{i+1}$ is deployed to filter the incoming requests. To meet  the real-time requirement,  instead of applying the model on each request,  it is applied to the mixture to obtain an updated  blacklist of attack  identities (e.g., IP addresses).  The blacklist is then activated to filter out the attack requests.   
 \end{itemize}
In the offline setting, the training can utilize the full $\mathcal{M}$ and $\mathcal{N}$, and there is no real-time requirement. 

\paragraph{\bf DDoS in the application layer.} 
A request in network-layer DDoS such as Syn flood typically consists of a single packet. In contrast,  a request in application-layer DDoS, which is the focus of this paper, could be a sequence $\bm{x} = (x_1, x_2, \ldots, x_{T})$ of sub-requests $x$'s.  For instance, a visit to a web-service could consist of an \texttt{HTTP-GET} to the landing page, followed by others.  The intrinsic characteristic of such a sequence could be learned to differentiate attacks.    In our notation, we view each request $\bm{x}$ as a sequence of sub-requests, and a user corresponds to an IP-address.   In evaluating performance, instead of measuring the actual byte counts,    the reported measures are calculated based on evaluation metrics such as accuracy, false-positive rate, F1 scores.


%% file: section/sec4-associate.tex
\section{N-over-D Framework} 
\label{sec:pan} 
We propose the N-over-D framework to address the formulated learning problem, extending a one-class unsupervised learning anomaly detection into the two-class classifier in a three-step process. A one-class learning model is trained on a training set containing samples of the same class, giving a predictive model that, on input $\bm{x}$, output the probability of having ${\bm x}$ in that class.
\begin{enumerate}
\item During pre-processing, one-class learning is applied to $\mathcal{N}$ to obtain a model N that predicts the distribution $P_n$. 
\item During online learning, a similar process is applied to $\mathcal{M}$ to obtain a model $D$ that predicts $P_d$.  
\item During filtering, the predicted probabilities, $P_n(\bm{x})$ and $P_d(\bm{x})$, are combined to give a model that predicts  $P_a(\bm{x})$, the likelihood that $\bm{x}$ is from the attack given that $\bm{x}$ is observed in the mixture. 
\end{enumerate}

The last step calculates $P_a(\bm{x})$ from $P_n(\bm{x})$ and $P_d(\bm{x})$,  following similar techniques in PacketScore~\cite{1354679}. Note that, 

\begin{equation} \label{eq:atl}
\begin{split}
   P_a(\bm{x}) & =  1-  P(\bm{x} \text{ is normal}\mid \text{request } \bm{x}  \text{ is observed in mixture}) \\ 
                       &  = 1- \frac{ (1-\alpha) P_n(\bm{x})}{\alpha P_d(\bm{x}) + (1-\alpha) P_n(\bm{x})} \\
    & = \frac{1}{1+  (\frac{1-\alpha}{\alpha})\frac{P_n}{P_d}(\bm{x})} \\
    \end{split}
\end{equation}  
From $P_a(\cdot)$,  we can classify $\bm{x}$ as an attack when $P_a(\bm{x})>0.5$. However, the accuracy would rely on the accuracy of the estimated $\alpha$.   
Typically, during filtering, a threshold is set so that a certain predetermined percentage of requests would be allowed. It is tolerable to accept attack requests as long as the system can handle the load. 
Such a threshold can be selected by sorting probabilities $P_a(\cdot)$ of the received requests and determining the appropriate cutoff.  Since only the ranking matters,  we can ignore the constant terms and focus only on the ratio $P_n/P_d$, which does not depend on $\alpha$. 
Moreover, to facilitate fast online learning, we could deploy transfer learning from $N$ to $D$ so that we can train $D$ readily. We design model $D$ to achieve quick online updates. When an attack occurs, the optimized embedding of $N$ is transferred to $D$, thereby significantly reducing the training needed from a random initialization of $D$.

\textit{Choice of one-class classification.}
Not all one-class classification is applicable in the N-over-D approach.   The candidate should meet the following two requirements.   (1) Firstly, Equation (\ref{eq:atl}) assumes that the output score of the model is the predicted probability.  Nevertheless, many models transform the probability and output a score that preserves the ranking but distorted the actual value.  The ratio of such distorted values from the two models might not preserve the ratio ranking.   Hence, it is essential that the score should be the probability, or with some additive and multiplicative constants.  (2) Secondly, as indicated earlier,  we need a mechanism to transfer learning from $N$ to $D$.

\textit{Limitations.}
The N-over-D approach has a limitation in that it relies on the accuracy of the one-class classification, and the classification is applied to the normal and mixture almost independently (except some information that is transferred from $N$ to $D$ via transfer learning). Intuitively, more accurate results could be attained via joint-learning. For instance, some combination of features values could have few samples from ${\mathcal N}$ but a large population from ${\mathcal M}$. When applied independently, $N$ might overgeneralize those features and predict a higher probability. However, with the additional information that there is indeed a high population from ${\mathcal M}$, a more refined and accurate boundary could be learned.

%% file: section/sec5-iterative.tex
\section{Iterative Classifier Framework} 
\label{sec:iterative}

In this section, to address the limitations of the N-over-D nets, we propose a method that optimizes a specifically formulated loss function designed for deep learning, which directly trains on both ${\mathcal N}$ and ${\mathcal M}$. 
We employ a binary classification model, similar to the model architecture of $N$ and $D$, and iteratively update it. Let the trained classifier $C_i$ be the model obtained after the $i$-th iteration.
\begin{enumerate}
\item At each interval $\ell$ during an attack, we assume all traffic is malicious.
\item We then train classifier $C_i$ on the newly observed data to update the model $C_{i+1}$.
\item Using classifier $C_{i+1}$, we predict and rank the observed data, selecting the top predicted attacks to retrain the classifier; by performing this procedure iteratively, the classifier directly minimizes our defined loss function (Equation \ref{eq:binaryloss}).

\end{enumerate}

Following the goal in the two-class learning problem (\cref{sec:formulate}), we define the following loss function.
\begin{equation} \label{eq:binaryloss}
\begin{split}
  \mathcal{L} = 
    &-   \frac{1}{|\mathcal N|}  \sum_{y\in {\mathcal N}}
    \log(1-p(y))   \\
  &-  \frac{1}{\alpha|\mathcal M|} 
            \sum_{y\in {Sel}_\alpha ({\mathcal M})}   \log(p(y))  \\
     &-   \frac{1}{(1-\alpha)|{\mathcal M}|}   \sum_{y\in {\mathcal M}, y\not\in{Sel}_\alpha ({\mathcal M})}   \log(1-p(y))  
\end{split}
\end{equation}   
where  $p(y)$ is the model-predicted probability of the observed input $y$ being the attack, and $Sel_\alpha({\mathcal M})$ contains $\alpha |{\mathcal M}|$ samples from ${\mathcal M}$ with the highest model-predicted probability
$p(\cdot)$.  The set $Sel_\alpha({\mathcal M})$ can be obtained by  ranking the samples in ${\mathcal M}$ according to the predicted probabilities and selecting the highest $\alpha|{\mathcal M}|$.  
In other words,  the three summations correspond to the three sets ${\mathcal N}$,  $Sel_\alpha({\mathcal M})$, and $({\mathcal M}-Sel_\alpha({\mathcal M}))$.  The loss function favors the first and third set to be classified as ``normal'',  and the second set to be classified as ``attack''.

The iterative classifier is able to perform fast online learning because it uses one supervised learning model, and it is suitable for attacks with a short timeframe. We assume all requests during the normal-day traffic as legitimate and all the traffic during an attack period, consisting of a mixture of both normal and attack traffic, as malicious and label them accordingly. By making such assumptions, we construct the initial set of inexactly labeled training data.

Next, we train the classifier using these inexact labels. We then predict the same initial training data set with this trained classifier, retain the top $40\%$ of traffic predicted as malicious to form a reduced attack dataset, and construct the next set of training data. By iteratively performing this prediction and reducing attack data, it allows the iterative classifier to repeatedly improve its model performance by increasingly retaining the traffic during an attack that is most likely to be truly malicious.

Furthermore, we design another classification model that trains on the entire training data of each attack using actual ground truth information. In this setting, we suppose that all normal and attack data is available during model training and that the labels accurately distinguish the attack from the normal traffic. The full classifier uses actual ground truth labels effectively provides us a soft performance upper bound. By taking a classification approach that leverages data label information, we can (1) study how this extra information boosts our approach if labels are available, and (2) construct powerful comparison methods.

%% file: section/sec6-lstm_based.tex
\section{LSTM N-over-D} 
\label{sec:associate}
    \begin{figure}[htbp]
        \centering
        \includegraphics[width=\linewidth]{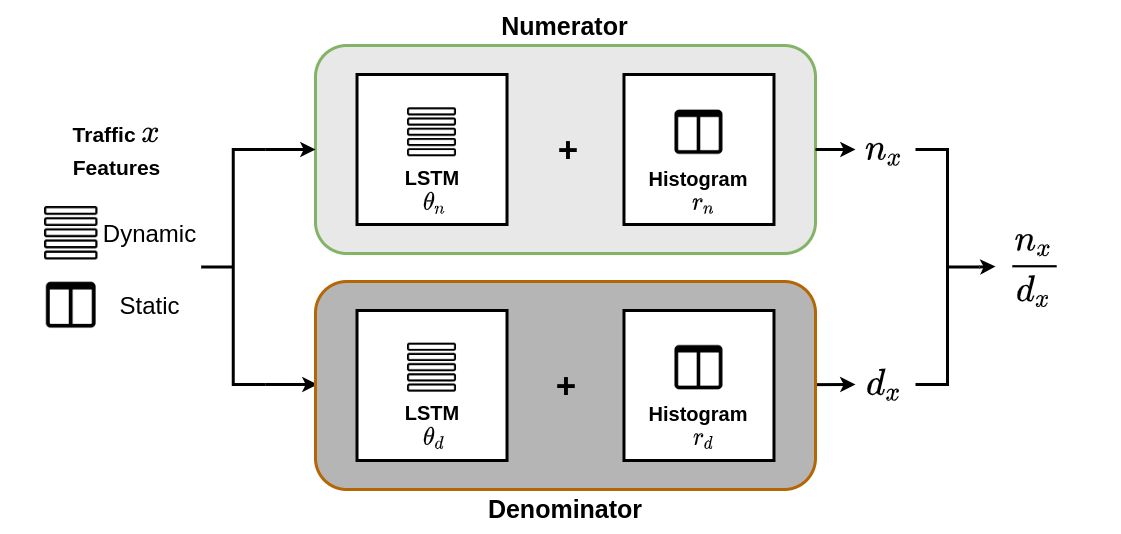}
        \caption{An overview of the LSTM-based N-over-D.}
        \label{fig:alliednets}
    \end{figure}  

The N-over-D nets (Figure~\ref{fig:alliednets}) takes a coordinated predictive approach that leverages sequential modeling and online learning to discriminate between normal and DDoS traffic. Our approach is inspired by the classic statistical DDoS mitigation work~\cite{1354679} that assigns each network request a score based on both probabilities that it is legitimate and malicious, which captures the differences in score distribution of attack and legitimate requests. N-over-D consist of two ``partners''---a normal-day model $N$ and a model $D$ for the DDoS duration. The goal of $N$ is to capture the distribution over the request $\bm{x}$ during normal network traffic conditions and predict the likelihood $n_x$ that the traffic is legitimate. At the same time, $D$ estimates the probability $d_x$ that each network unit of data is a malicious DDoS attack. Given incoming traffic, $N$ and $D$ evaluate the traffic for normality and abnormality respectively and make a joint decision $n_x/d_x$ on the legitimacy of the traffic. We provide details of the input features, model architectures, and design choices below (\cref{subsec:features,subsec:numerator,subsec:denominator}).

\subsection{Sequence Data Preparation} 
\label{subsec:features}

\begin{table*}[tbp]
  \centering
  \caption{Details of the eight dynamic and two static features in a sequence sample.}
\resizebox{\linewidth}{!}{%
\begin{tabular}{|c|c|c|c|c|c|c|c|c|c|c|c|c|}
\hline
\multicolumn{1}{|c|}{\textbf{Req \#}} & \multicolumn{1}{c|}{\textbf{\begin{tabular}[c]{@{}c@{}}Absolute\\ Time\end{tabular}}} & \multicolumn{1}{c|}{\textbf{\begin{tabular}[c]{@{}c@{}}Request\\ Len\end{tabular}}} & \multicolumn{1}{c|}{\textbf{\begin{tabular}[c]{@{}c@{}}IP\\ Flags\end{tabular}}} & \multicolumn{1}{c|}{\textbf{\begin{tabular}[c]{@{}c@{}}TCP\\ Len\end{tabular}}} & \multicolumn{1}{c|}{\textbf{\begin{tabular}[c]{@{}c@{}}TCP\\ Ack\end{tabular}}} & \multicolumn{1}{c|}{\textbf{\begin{tabular}[c]{@{}c@{}}TCP\\ Flags\end{tabular}}} & \multicolumn{1}{c|}{\textbf{\begin{tabular}[c]{@{}c@{}}TCP\\ Window Size\end{tabular}}} & 
\multicolumn{1}{c|}{\textbf{\begin{tabular}[c]{@{}c@{}}Highest Layer\\ (Protocol)\end{tabular}}} & \multicolumn{1}{c|}{\textbf{\begin{tabular}[c]{@{}c@{}}Extra\\ Info\end{tabular}}} & \multicolumn{1}{c|}{\textbf{\begin{tabular}[c]{@{}c@{}}List of\\ Protocols\end{tabular}}} \\ \hline \hline
1   &23352.0   &66.0   &0x00004000   &0.0   &19.0   &0x00000012   &14600.0   &TCP  &\multirow{8}{*}{\begin{tabular}{p{2cm}p{c}} \\ 443 $\to$ 8918 [SYN, ACK] Seq=0 Ack=1 Win=14600 Len=0 MSS=1460 SACK\_PERM=1 WS=512 \end{tabular}}   &\multirow{8}{*}{eth:ethertype:ip:tcp}   \\ \cline{1-9} 
2   &23352.1   &60.0   &0x00004000   &0.0   &30.0   &0x00000010   &15872.0   &TCP   &   &   \\ \cline{1-9}
\vdots  &\vdots   &\vdots   &\vdots   &\vdots   &\vdots   &\vdots   &\vdots   &\vdots   &     &   \\ \cline{1-9}
$t-1$   &23368.5   &85.0   &0x00004000   &31.0   &42.0   &0x00000018   &16896.0   &TLSv1.2   &   &   \\ \cline{1-9} 
$t$     &23368.7   &60.0   &0x00004000   &0.0   &42.0   &0x00000011   &16896.0   &TCP   &   &   \\ \cline{1-9}
$t+1$   &0   &0   &0   &0   &0   &0   &0   &0   &   &   \\ \cline{1-9}
\vdots  &\vdots   &\vdots   &\vdots   &\vdots   &\vdots   &\vdots   &\vdots   &\vdots   &   &   \\ \cline{1-9}
$T$     &0   &0   &0   &0   &0   &0   &0   &0   &   &   \\ \hline
\end{tabular}
}
\label{tab:features}
\end{table*}

The data processing procedure for online learning widely differs from that of a static dataset of requests. Hence, we prepare the data into a streaming form; sequences of sub-requests, that are grouped by source and destination IP addresses.

Let $\tau_0$ be the starting interval of the traffic and $\tau_i$ are the following $\ell$-min intervals, where $\tau_1$ is the first interval, $\tau_2$ the second, and so forth. By extracting the intervals of network traffic into arrays, we prepare the data for sequential online learning, which allows the LSTM to learn the distribution of both normal and DDoS attack traffic. 
In the rest of this section, we describe in detail each stage of the preparation process (see Appendix for details of the data preprocessing procedure algorithm).

During each interval $\tau_i$, the preparation procedure captures network traffic that is collected for an allocated amount of time $\ell$. The request sequences are then constructed using the sub-requests gathered during the interval. 
We give an example of a sequence of sub-requests (\cref{tab:features}).
The example sequence consists of $T$ sub-requests with attributes, consisting of both dynamic and static attributes. 
In our case, we chose eight dynamic attributes that capture the information that varies from request to request and two static attributes that are most common for all requests for a sequence. 
Sequences longer than $T$ are split to form a new sequence, while sequences shorter than $T$ are zero-padded at the end. 

\subsection{Model for Normal Traffic}
\label{subsec:numerator}
The model $N$ for normal-day traffic is made up of two particular components: (1) a probabilistic sequential learning model that learns to predict sequences of temporal features, and (2) a histogram frequency distribution approximate representation to model the static features extracted from the network traffic. We model $N$ using a parameterized LSTM network $\bm{\theta}_n$ and a Histogram method $\bm{r}_n$. 

\begin{figure}[htbp]
        \centering
        \includegraphics[width=\linewidth]{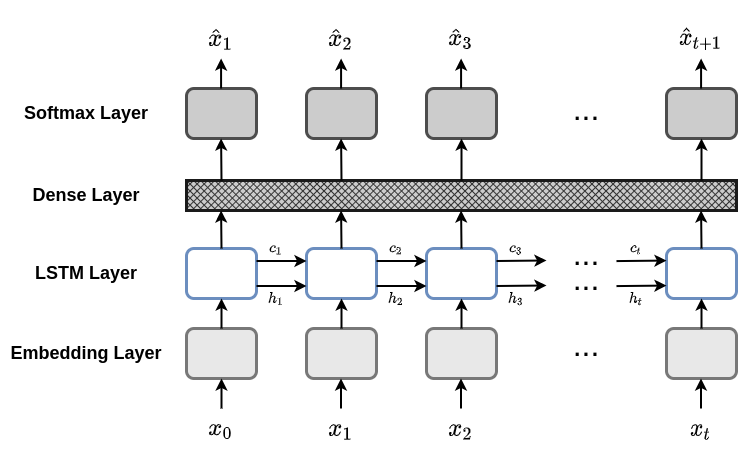}
        \caption{The LSTM model architecture.}
        \label{fig:lstm}
    \end{figure}

\textbf{LSTM.}
At each interval $\tau_i$, the LSTM $\bm{\theta}_n$ takes as input the sequence of requests $\bm{x} = (x_1, x_2, \ldots, x_{T})$ that are the temporal features in the sequences. We set $T = 200$, balancing between efficiency of the model and memory requirements to learn the data. The LSTM learns from inputs at each step to output the predictions $(\hat{x}_1, \hat{x}_2, \dots, \hat{x}_{T})$; the training process is illustrated in Figure~\ref{fig:lstm}, and the hyperparameters are detailed below.  

\textit{Input layer}: Recall in \cref{subsec:features} that we preprocess the requests into intervals of sequences, simulating a real-time data streaming process. By creating an environment suitable for sequential online learning, it enables the LSTM to leverage its modeling capabilities to learn the correlation between traffic requests in the same sequence. This layer takes as inputs the data sequences $\bm{x}$ of each interval $\tau_i$, represented as a 3D tensor $T \times m \times l$, where $T$ is the length of the sequences and each $\bm{x}_t$ is represented as a $m \times l$ matrix. The matrix contains $m$ samples and $l$ features.

\textit{Embedding layer}: The inputs $\bm{x}$ are then passed to the embedding layer $emb_n$, which creates a dense matrix that is learned through backpropagation of the training process. This layer reduces all features into a dense vector representation of dimension 16. The embedding, a dense matrix in a linear space, achieves two important functions. First, by using an embedding with a much smaller dimension than the feature size, it reduces the dimension of feature representations to the embedding size and reduce the model complexity. Second, the input feature embedding groups semantically similar features in an Euclidean space, and allow for a better feature representations.

\textit{LSTM layer}: The outputs from the embedding layer then form the compact inputs to the LSTM layer, which is a powerful machine learning architecture to learn from sequence data. LSTMs are designed to predict the presented observations at each timestep sequentially. 
By taking a sequence $\{x_1, x_2, \ldots, x_T\}$ as input, the LSTM constructs a corresponding sequence of hidden state representations $\{h_1, h_2, \dots, h_T\}$. A single-layer LSTM uses the hidden representations $\{h_1, h_2, \dots, h_T\}$ for estimation and prediction. In our case, we use two LSTM layers, where the second layer uses the hidden states of the previous layer as inputs. Every hidden state in each layer performs memory-based learning to remember relevant features using previous inputs. Previous hidden states and current inputs are transformed into a new hidden state, and it is achieved through a recurrent operator that takes in $(h_{t-1}, x_t)$, such as:

\begin{equation}\label{eq:recurrentop}
  h_t = \tanh(W_h h_{t-1} + W_x x_t + b)
\end{equation}
where $W_h$, $W_x$, and $b$ are parameters of the layer and $\tanh(\cdot)$ represents the standard hyperbolic tangent function. 

The LSTM is made up of these cells that are specifically designed for series data to trace the history from previous network requests. In order to overcome both the vanishing gradient and long-term dependency issues, each LSTM cell retains an internal cell state $c_t$ and a hidden state that is the output. Both $c_t$ and $h_t$ are computed via three gate functions to retain both long and short term storage of information. The forget gate $f_t$, input gate $i_t$, and output gate $o_t$ control the flow of information.
In this work, we use the LSTM without peep-hole connections to handle complex sequences with long-range structure. 
We design each cell to consist of 32 units, the dimensionality of the output space, which is passed to the prediction layer (see Appendix for details).

\textit{Prediction layer}: Following the process (Figure~\ref{fig:lstm}), the LSTM takes in $x_t$ at each step $t$ and makes a prediction $\hat{x}_{t}$ of the request at the next step $t+1$. We set the output function as the softmax function:

\begin{equation} \label{eq:softmax}
    \text{Softmax}(x_{t}) = \frac{\exp(x_t)}{\sum_j \exp(x_j)}
\end{equation}   
where $j$ is the total number of possible outputs. The softmax function converts a real vector to a vector of categorical probabilities, where the elements of the output vector are in range $(0, 1)$ and sum to one. We can interpret the results as a probability distribution and take the maximum as the prediction $P(x_{t+1} \mid x_{0}, \dots, x_{t})$ of the next step in the sequence, where $x_{0}$ is the start of sequence symbol. We start all sequences with a this symbol and the probability of observing it at step $0$ is one.

Hence, by assuming that the state of each time step depends on previous time steps where the current information has a dependency on previous data, we can use standard conditional probability theory to determine the joint probability of each sequence of requests by calculating:

\begin{equation} \label{eq:lstmpredict}
\begin{split}
    P(\bm{x}) & = P(x_{0}) \cdot P(x_{1} \mid x_{0}) \cdots P(x_{T} \mid x_{0}, \dots, x_{T-1}) \\
    & = P(x_{0}, x_{1}, \dots, x_{T})
    \end{split}
\end{equation}   
and assigning this computed probability as the \textit{LSTM prediction} $\bm{\theta}_n (\bm{x})$.

When training the LSTM network $\bm{\theta}_n$, our objective is to minimize the difference between the model prediction and observed request at each timestep $t$. The loss function we minimize is defined as:

\begin{equation} \label{eq:lstmloss}
    \mathcal{L} = - \sum_{c=1}^{M} y_{t,c} \log(\hat{x}_{t,c})    
\end{equation}   
where $M$ is the total number of possible next request, $y_{t,c}$ is the indicator if ground-truth label $c$ is the correct prediction for observation $x_{t,c}$ in the $t$-th step of the sequence, and $\hat{x}_{t,c}$ is the LSTM predicted probability of the input at time $t$. As the possibilities $M$ can be exceedingly large, it could cause an explosive increase in model computational complexity whereby drastically increasing the time required to train the model. A large $M$ presents a potential model robustness issue in large datasets. Moreover, there could be other considerations, such as memory limitations and GPU computational constraints. Hence, we employ a feature hashing trick~\cite{10.1145/1553374.1553516} to reduce $M$ and solve the mentioned issues of a large number of request possibilities.


\textbf{Histogram.}
For each sequence of unique source and destination IP addresses, the static features are inputs for the histogram model $\bm{r}_n$. The \texttt{Extra Info} and \texttt{List of Protocols} attributes are the responses of the server after receiving requests and the list of network protocols used. They generally remain constant throughout the entire sequence. Since the two features are observed to be static, we assume independence between them and the dynamic features.

A histogram model is used to capture the approximate frequency distribution of the static attributes. We count the number of specific \texttt{Extra Info} and \texttt{List of Protocols} pairs, and calculate the histogram probability prediction of each pair:

\begin{equation} \label{eq:nhist_predict}
\begin{split}
    \bm{r}_n(\bm{x}) & = \frac{m_i(\bm{x})}{\sum_{i=1}^{k} m_i}
\end{split}
\end{equation} 
where $\bm{x}$ is a static feature pair, $m_i$ is a function that counts the number of observations of a particular pair, and $k$ is the total number of static feature pairs.

\textbf{Model $N$ prediction.}
To predict how likely a sequence belongs in the normal traffic, we incorporate both the \textit{LSTM prediction} $\bm{\theta}_n (\bm{x})$ and the \textit{Histogram prediction} $\bm{r}_n(\bm{x})$. By taking the product of the two predictions using dynamic features and static features: 

\begin{equation} \label{eq:npredict}
\begin{split}
    P_n(\bm{x}) & = \text{\textit{$N$ LSTM}} (\bm{x}) \times \text{\textit{$N$ Hist.}} (\bm{x})  \\
              & = \bm{\theta}_n(\bm{x}) \times \bm{r}_n(\bm{x}) \\
\end{split}
\end{equation}   
we obtain the probability of a sequence of network requests, which serves as the numerator of the N-over-D framework. 

\subsection{Model for Mixture Traffic}
\label{subsec:denominator}
The model $D$, similar to the normal-day model $N$, also consists of a parameterized LSTM and a Histogram model that learns to predict both sequences of dynamic features and static features of the traffic, respectively. The goal of $D$ is to learn meaningful generalization of the mixture traffic during DDoS to predict the probability of each network request to be malicious. It is designed to perform fast online learning, updating the models on the evolving attack traffic every minute. 

The model architecture is designed with a significant difference to promote the online learning design. We initialize the embedding layer $emb_d$ with the previously trained embedding $emb_n$ during normal traffic. Since the embedding is a compact representation of the input network requests, we considerably minimize the training of $\bm{\theta}_d$ by transferring the learned embedding layer rather than initializing $emb_d$ randomly. 
It significantly reduces the number of model parameters, which decreases the computational cost, to achieve fast learning and inference during attacks. 
In addition, we employ a similar Histogram model $\bm{r}_d$ on the static features of network requests during the DDoS attacks.

\textbf{Model $D$ prediction.}
To predict how likely a sequence belongs in the malicious traffic, we incorporate both the $D$ \textit{LSTM prediction} $\bm{\theta}_d (\bm{x})$ and the $D$ \textit{Histogram prediction} $\bm{r}_d(\bm{x})$. By taking the product of the two predictions using dynamic features and static features: 

\begin{equation} \label{eq:dpredict}
\begin{split}
    P_d(\bm{x}) 
            & = \bm{\theta}_d(\bm{x}) \times \bm{r}_d(\bm{x}) \\
\end{split}
\end{equation}
we obtain the probability of a sequence of network requests during DDoS attacks, which serves as the denominator of the N-over-D framework.

\subsection{N-over-D Likelihood Measure}
\label{subsec:defense}
N-over-D performs by first learning $N$ to model the normal traffic data distribution and predict the likelihood $P_n(\bm{x})$ that the presented network request sequences are legitimate. Next, during DDoS attacks, the online-learner $D$ is trained for accelerated prediction $P_d(\bm{x})$ of the incoming DDoS traffic to estimate the probability that each sequence is malicious. Finally, we adopt an integrated approach to decide the normality of a sequence by taking the division:

\begin{equation} \label{eq:alliednets}
\begin{split}
    \frac{N}{D}(\bm{x}) & = \log\Bigg( \frac{P_n(\bm{x})}{P_d(\bm{x})} \Bigg) \\
\end{split}
\end{equation} 
Since a high $P_n(\bm{x})$ indicates high legitimacy and a low $P_d(\bm{x})$ suggests a small chance of maliciousness, the higher the joint probabilities $\frac{N}{D}(\bm{x})$, the more likely the network traffic is normal. Hence, we differentiate between normal and attack traffic by ranking the combined probabilities of each sequence in the DDoS traffic of each $1$-min interval, and discarding request sequences with relatively low joint probabilities.

%% file: section/sec7-eval.tex
\section{Experiments}
\label{sec:experiments}
In this section, we present the evaluation of our approaches performed on various DDoS attacks. We demonstrate the effectiveness of the models and compare them with state-of-the-art methods.


\subsection{Datasets}
\label{subsec:datasets}
To validate the N-over-D and Iterative Classifier, we use three separate volumetric DDoS attacks from two publicly available network datasets, (1) CICIDS2017~\cite{icissp18} dataset~\footnote{www.unb.ca/cic/datasets/ids-2017.html}, which contains generated traffic resembling real-world attacks, and (2) CAIDA UCSD ``DDoS Attack 2007''~\cite{caida07} dataset~\footnote{www.caida.org/data/passive/ddos-20070804\textunderscore dataset.xml} that contains anonymized traffic traces from a DDoS attack on August 4, 2007. 
A summary (\cref{tab:benchmarkdata}) of the test data during attacks, shows the actual proportion $\alpha$ of attack traffic. 



\begin{table}[htbp]
  \centering
  \caption{Test data samples. Both \textit{HULK} and \textit{LOIC} attacks are from the \textit{CICIDS2017} benchmark~\cite{icissp18}, and the 2007 Attack is from \textit{CAIDA07} real-world~\cite{caida07} dataset.}
\resizebox{.9\linewidth}{!}{%
\begin{tabular}{|c|c|c|c|c|}
\hline
\textbf{Dataset} & \textbf{\# Total} &\textbf{\# Normal} & \textbf{\# DDoS} & \textbf{DDoS (\%)} \\ \hline
\textit{HULK}      &11710   &4552   &7158   &61.1   \\ \hline
\textit{LOIC}     &2447   &1069   &1378   &56.3   \\ \hline
\textit{CAIDA07}     &16310   &3262   &13048   &80.0   \\ \hline
\end{tabular}
}
\label{tab:benchmarkdata}
\end{table}

\subsubsection{Benchmark Data}
\label{subsec:benchmark}
We evaluate our approach with one of the latest benchmark data \textit{CICIDS2017}~\cite{icissp18}. It is a synthetically generated dataset produced by the Canadian Institute for Cybersecurity of the University of New Brunswick (UNB), Canada. 
The labeled data consists of the most up-to-date common attacks of network activity, including two volumetric attacks, one on Wednesday and one on Friday. The traffic traces provided in the \textit{pcap} format include full packet payloads, data labels, and traffic flow statistics. The types of attacks and the times they occur are listed below: 

\begin{enumerate}
\setlength\itemsep{0em}
    \item HULK (Wed 10:43 a.m. -- 11 a.m.) 
    \item Low Orbit Ion Canon (LOIC) (Fri 3:56 p.m. -- 4:16 p.m.) 
\end{enumerate}
The attacks last for 17 and 20 mins, respectively. In order to obtain normal traffic to learn the meaningful generalization of legitimate users, we retrieve traffic data prior to the attacks. For attack (1), we use normal network traffic on Wednesday (9:00 a.m. -- 9:47 a.m.), and for attack (2), we take Friday (3:00 p.m. -- 3:56 p.m.) data to train the model $N$. In each attack, after training $N$ on the respective normal traffic, we perform online learning and detection using $D$ when an attack occurs, demonstrating progressive improvement of the detection performance. We leave the last five intervals as the test set, to evaluate our models and compare with state-of-the-art methods.

\begin{figure*}[!ht]
\caption{The x-axis is the rejection ratio (0--1), and the y-axis is the corresponding false rejection rate. If the rejection ratio is at 0.8, it means that 80\% of the requests are rejected, and the graph shows the rate of false positives.} 
\centering
    \begin{minipage}[b]{.33\textwidth}
    \centering
    \subfloat[][HULK]{
        \includegraphics[width=1.05\linewidth]{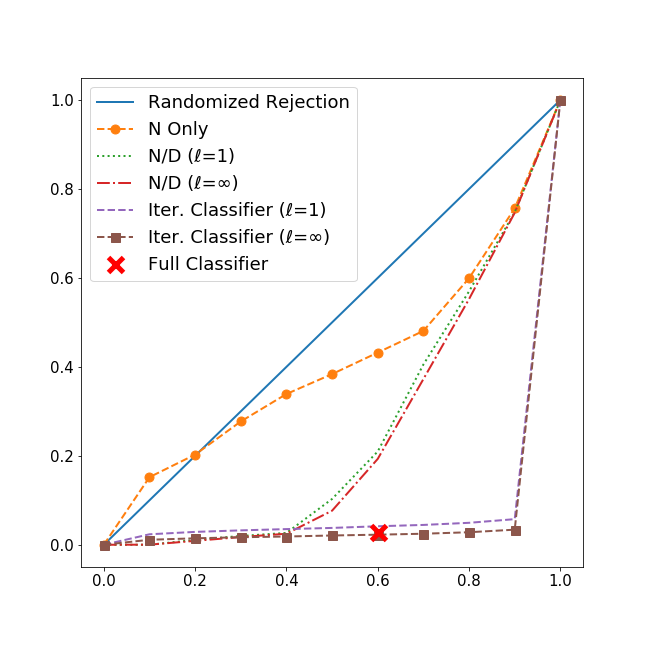}}
    \end{minipage} 
    \begin{minipage}[b]{.33\textwidth}
    \centering
    \subfloat[][LOIC]{
        \includegraphics[width=1.05\linewidth]{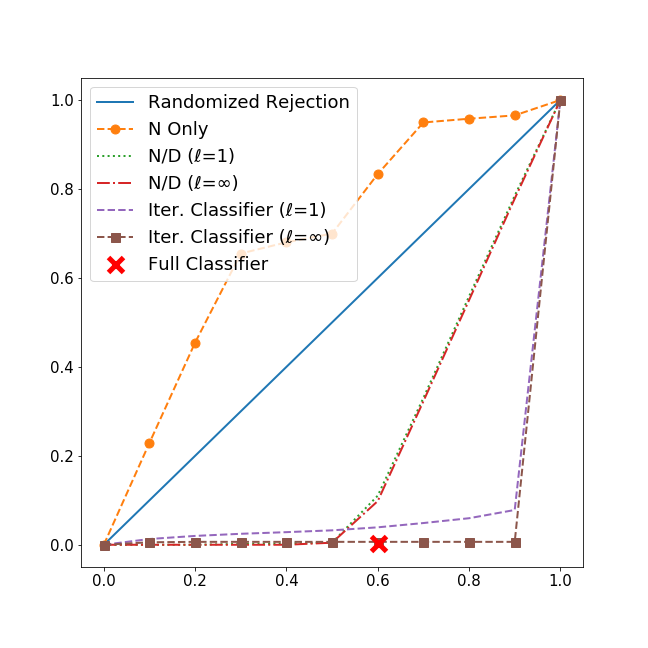}}
    \end{minipage} 
    \begin{minipage}[b]{.33\textwidth}
    \centering
        \subfloat[][CAIDA07]{
        \includegraphics[width=1.05\linewidth]{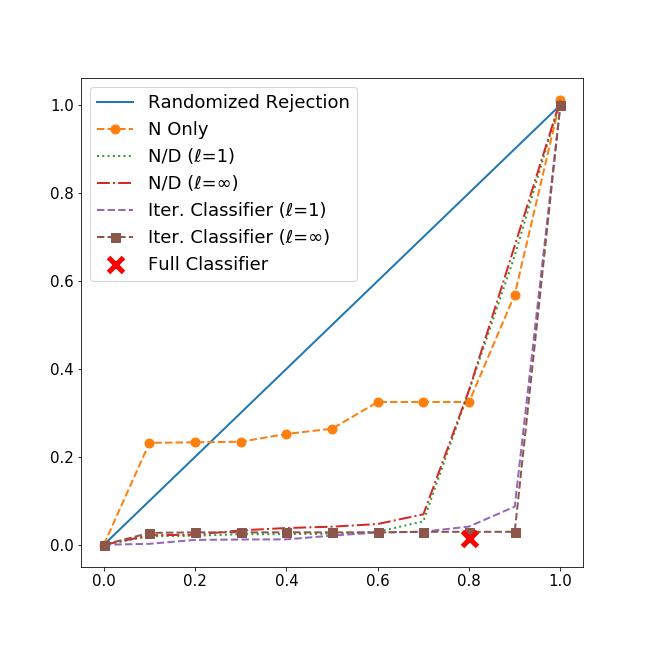}}
    \end{minipage} 
    
\label{fig:GGval_test}

\captionof{table}{Detection performance on the various attacks in \textit{CICIDS2017} and \textit{CAIDA07} datasets.}
\resizebox{0.6\linewidth}{!}{%
\begin{tabular}{c||c|ccccc}
\hline
\textbf{Dataset} & \textbf{Model} & \textbf{ACC} & \textbf{FPR} & \textbf{Prec.} & \textbf{Rec.} & \textbf{F1} \\ \hline 

\multirow{6}{*}{HULK} &N &0.5903 &0.3837 &0.7016 &0.5739 &0.6313   \\ 
&N-over-D ($\ell=1$)    &0.8086 &0.1030 &0.9198 &0.7524 &0.8277   \\
&N-over-D ($\ell=\infty$)   &0.8291 &0.0766 &0.9403 &0.7692 &0.8462   \\

&Iterative Classifier ($\ell=1$)  &0.9088	&0.0378	&0.8831	&0.7659	&0.8204 \\ 
&Iterative Classifier ($\ell=\infty$)   &0.9120	&0.0206	&0.9299	&0.7313	&0.8187 \\ \cline{2-7} 
&Full Classifier   &0.9350	&0.0281	&0.9808	&0.9116	&0.9449 \\ \hline

\multirow{6}{*}{LOIC} &N &0.3257 &0.6997 &0.3889 &0.3454 &0.3659   \\ 
&N-over-D ($\ell=1$)   &0.9330 &0.0047 &0.9959	&0.8846	&0.9370   \\ 
&N-over-D ($\ell=\infty$)   &0.9330	&0.0047	&0.9959	&0.8846	&0.9370   \\ 
&Iterative Classifier ($\ell=1$)   &0.8805	&0.0329	&0.8808	&0.6494	&0.7476 \\ 
&Iterative Classifier ($\ell=\infty$)   &0.9486	&0.0220	&0.9016	&0.8275	&0.8630 \\ \cline{2-7}
&Full Classifier   &0.9713 &0.0037 &0.9968 &0.9121 &0.9526   \\ \hline

\multirow{6}{*}{CAIDA07} &N &0.6009	&0.2639	&0.8958	&0.5671	&0.6945   \\ 
&N-over-D ($\ell=1$)   &0.6944	&0.0251	&0.9900	&0.6242	&0.7657   \\
&N-over-D ($\ell=\infty$)   &0.6880	&0.0411	&0.9837	&0.6202	&0.7608   \\ 

&Iterative Classifier ($\ell=1$)   &0.7081	&0.0208	&0.9919	&0.6403	&0.7783 \\ 
&Iterative Classifier ($\ell=\infty$)   &0.7088	&0.0141	&0.9945	&0.6396	&0.7785 \\ \cline{2-7}
&Full Classifier   &0.7094	&0.0120	&0.9953	&0.6398	&0.7789   \\ \hline 

\end{tabular}
}
\label{tab:benchmarkperf}
\end{figure*}

\subsubsection{Real-world Data}
\label{subsec:realworld}
The other dataset, \textit{CAIDA07} UCSD ``DDoS Attack 2007''~\cite{caida07}, is a real-world attack collected from various locations worldwide. It occurred on 04 August, 2007. The data contains approximately one hour (20:50 UTC to 21:56 UTC) of anonymized traffic traces from a DDoS attack that includes types of network traffic such as Web, FTP, and Ping. 
We use all traffic traces before the attack (20:50 -- 21:12) as normal network traffic to train the model $N$. When the attack starts at 21:13, we perform both online and offline learning of $D$ on the first 4 mins; the next min of the attack is left as the test set to evaluate our models. The summary (\cref{tab:benchmarkdata}) of the \textit{CAIDA07} test data shows $80\%$ attack traffic.

\subsection{Evaluation Results}
We present a detailed evaluation of the proposed approaches on both the \textit{CICIDS2017} and \textit{CAIDA07} datasets.
Setting the rejection rate at $50\%$ and online interval $\ell$ at 1-min, we measure the performance of iterative classifier ($\ell=1$) and N-over-D ($\ell=1$). We also set the interval ($\ell=\infty$) as all the data during attack for both approaches, and present the Accuracy (ACC), False Positive Rate (FPR), Precision (Prec.), Recall (Rec.), and F1 Score (F1) results. Figure~\ref{fig:GGval_test} and \cref{tab:benchmarkperf} illustrate these results. As demonstrated, both N-over-D and iterative classifier achieve low error rates and high accuracy scores across various attacks. These results validate the robust design of our approach.

In \cref{tab:benchmarkperf}, the results demonstrate that the iterative classifier, which directly optimizes the loss function (Equation \ref{eq:binaryloss}), achieves better performance than N-over-D, which estimates conditional probabilities of the normal-day and mixture traffic distributions. The iterative classifier is also more robust in FP rates across various rejection ratios (0--1), demonstrated in Figure~\ref{fig:GGval_test}. However, its empirical runtime (\cref{tab:runtime}) is roughly three to five times slower than that of N-over-D.

\begin{table}[htbp]
  \centering
  \caption{Empirical runtimes per epoch ($\ell=1$) for N-over-D and Iterative Classifier in seconds.}
\resizebox{.75\linewidth}{!}{%
\begin{tabular}{|c|c|c|}
\hline
\textbf{Dataset} & \textbf{N-over-D (s)} &\textbf{Iterative Classifier (s)}  \\ \hline
\textit{HULK}       &8      &41        \\ \hline
\textit{LOIC}       &3      &8          \\ \hline
\textit{CAIDA07}    &9      &43      \\ \hline
\end{tabular}
}
\label{tab:runtime}
\end{table}

Moreover, the iterative classifier is sensitive to the estimated proportion of attack $\alpha$ in the loss function, whereas N-over-D is not. In our experiments, we use $\alpha=0.6$ as a standard, but in other attacks where the attack proportion could be much higher, setting an appropriate $\alpha$ is crucial to the detection performance. 

Reported evaluation metrics are Accuracy (ACC), False Positive Rate (FPR), Precision (Prec.), Recall (Rec.), and F1 Score (F1). 
The evaluation metrics are defined as:

\begin{itemize}
\setlength\itemsep{0em}
    \item[] ACC = $(TP + TN) / (TP + TN + FP + FN)$ 
    \item[] FPR = $FP / (FP + TN)$ 
    \item[] Prec. = $TP / (TP + FP)$ 
    \item[] Rec. = $TP / (TP + FN)$ 
    \item[] F1 = $2 [(Prec. \times Rec.) / (Prec. + Rec.)] $
\end{itemize}
where $TP=\text{True Positives}$, $TN=\text{True Negatives}$, $FP=\text{False Positives}$, $FN=\text{False Negatives}$. We run all the experiments on NVIDIA Tesla P100 GPUs with 12 GB memory. The models have been implemented in Python v3.7.5 using the Keras v2.2.4 library on top of the Tensorflow v1.14.0 machine learning framework.

\subsection{Result Comparison}
In addition, we compare our approaches, N-over-D and Iterative Classifier, with state-of-the-art DDoS detection methods on the \textit{CICIDS2017} and \textit{CAIDA07} datasets. We note that the other studied methods leverage on the labels of the data for their classification models. However, as mentioned in the problem formulation, labeled data is a commodity that is often unavailable during the time of an attack.

\textbf{\textit{CICIDS2017}.}
We compare our approach with three online and four offline methods. 
The three online approaches include a CNN classifier LUCID~\cite{Doriguzzi_Corin_2020} that uses labels and an unsupervised method, Smart Detection~\cite{lima2019smart} that employs simple decision tree learning algorithms, and E-KOAD~\cite{DANESHGADEHCAKMAKCI2020102756}. 
A network flow graph-based detector DeepGFL~\cite{yao2018deepgfl}, and four deep-learning classifiers introduced by Cyber Security in IoT Networks~\cite{8666588} are offline methods. The four deep learning models are the MLP, 1D-CNN, LSTM, and 1D-CNN+LSTM  models. 

We note that the closest comparison, E-KOAD, presented the performance of the attack in the \textit{CICIDS2017} dataset that they achieved the best results. However, their smallest reported interval, $\ell$, is 2 mins, while we used 1-min intervals. Hence, we report the results for $\ell=2$ min too. 
We present the evaluation results of all the models and contrast it with our proposed models (\cref{tab:benchmarkcompare}).

\begin{table}[htbp]
  \centering
  \caption{Performance comparison with existing methods on the \textit{CICIDS2017} dataset. We note that while N-over-D and Iterative Classifier do not use label information for identifying attacks, almost every other compared method leverages these labels to achieve the reported results.}
  \resizebox{\linewidth}{!}{%
  \begin{tabular}{c||ccccccc}
    \hline
    \textbf{Method} & \textbf{ACC} & \textbf{FPR} & \textbf{Prec.} & \textbf{Rec.} & \textbf{F1} & \textbf{Labels} & \textbf{Online}\\ \hline 
    N-over-D  &0.936	&0.001	&0.999	&0.887	&0.940 &No &$\ell=2$ \\ 
    Iter. Classifier &0.935	&0.034	&0.852	&0.810	&0.831 &No &$\ell=2$ \\     

    Smart Detect~\cite{lima2019smart}   &0.800 &0.002 &0.992 &N/A &0.992 &No &$\ell=1.7$ \\
    E-KOAD~\cite{DANESHGADEHCAKMAKCI2020102756}   &0.995   &0.170   &N/A   &0.952   &N/A &No &$\ell=2$  \\         
    LUCID~\cite{Doriguzzi_Corin_2020}   &0.996   &0.005   &0.993   &0.999   &0.996 &Yes &$\ell=1.7$  \\ 
    DeepGFL~\cite{yao2018deepgfl}    &N/A   &N/A   &0.756   &0.302   &0.432 &Yes &No   \\ 
    MLP~\cite{8666588}   &0.863   &N/A   &0.884   &0.862   &0.873 &Yes &No   \\ 
    1D-CNN~\cite{8666588}   &0.951   &N/A   &0.981   &0.901   &0.939 &Yes &No   \\ 
    LSTM~\cite{8666588}   &0.962   &N/A   &0.984   &0.898   &0.895 &Yes &No   \\ 
    1D-CNN+LSTM~\cite{8666588}   &0.971   &N/A   &0.974   &0.991   &0.982 &Yes &No   \\ 
  \end{tabular}
}
\label{tab:benchmarkcompare}
\end{table}

The four learning models~\cite{8666588} are constructed by combining LSTM, CNN, and fully connected layers. Out of these architectures, designed for DDoS detection in Internet of Things (IoT) networks, the 1D-CNN+LSTM model performs the best. While it produces good classification results, the other models seem to suffer from low true-positive rates. As for DeepGFL~\cite{yao2018deepgfl}, it uses a graph representation of low-order features to perform classification. The results exhibit its weakness in identifying true positives, leading to low recall and F1 scores. Even though LUCID~\cite{Doriguzzi_Corin_2020}, a lightweight deep learning DDoS detection system, presents superior classification results, the model applies a CNN to network traffic flows to detect attacks that heavily rely on the accuracy of each traffic flow's labels to achieve such performance. 

On the other hand, N-over-D and Iterative Classifier achieve one of the highest scores in every evaluation metric while retaining low false-positive rates. It demonstrates competitive performance without utilizing the label information of the data.

\textbf{\textit{CAIDA07}.}
The compared methods include a self-organizing map detection scheme SOM~\cite{5735752}, a support vector machine classifier SVM~\cite{7562606}, a hybrid model SVM-SOM~\cite{7816865}, and an enhanced history-based IP filtering method eHIPF~\cite{8630919}. We present the evaluation results of all the models and contrast it with the N-over-D performance (\cref{tab:realworldcompare}).

\begin{table}[hbp]
  \centering
  \caption{Performance comparison with existing methods on the \textit{CAIDA07} real-world dataset. We note that while N-over-D does not use label information for identifying attacks, the compared methods consider these labels to achieve such results.}
  \resizebox{.9\linewidth}{!}{%
  \begin{tabular}{c||cccccc}
    \hline
    \textbf{Method} & \textbf{FPR} & \textbf{TPR} & \textbf{Rec.} & \textbf{F1} & \textbf{Labels} & \textbf{Online}\\ \hline 
    N-over-D           &0.025	&0.990	&0.624	&0.765 &No &$\ell=1$   \\ 
    Iter. Classifier   &0.020	&0.991	&0.640	&0.778 &No &$\ell=1$ \\ 
    SOM~\cite{5735752}   &0.065   &0.934   &N/A   &N/A &No &No  \\ 
    SVM~\cite{7562606}   &0.076   &0.923   &N/A   &N/A &Yes &No  \\ 
    SVM-SOM~\cite{7816865} &0.038   &0.981   &N/A   &N/A &Yes &No  \\ 
    eHIPF~\cite{8630919}   &0.006   &0.992   &N/A   &N/A &Yes &No  \\ 
  \end{tabular}
}
\label{tab:realworldcompare}
\end{table}

Methods~\cite{5735752,7562606,7816865} that adopt the SOM and SVM models produced relatively lower evaluation results, ranging from around 4\%--8\% false-positive rate and 92\%--98\% true-positive rate. As these are basic machine learning architectures, the models might not capture the high complexity of the network traffic as well as more complex neural architectures. Even though eHIPF~\cite{8630919} yields outstanding results, it is a system that invokes an intricate process that requires data labels. The performance of N-over-D, without the need for label information, proves to be very close to that of eHIPF.

\subsection{Analysis}
\label{sec:analysis}
While we show in the above results (\cref{tab:benchmarkperf}) that the N-over-D and Iterative Classifier are effective against a range of real-world and synthetic DDoS attacks, the contrast between traffic distributions during normal and DDoS periods for this effect has not been illustrated. 
We study the performance of various online parameters $\ell$, analyze these distributions from an empirical perspective, and discuss the pros and cons of the proposed approaches.


First, we present the performance (\cref{tab:lparams}) for the proposed online approaches based on various intervals $\ell$. It clearly shows that the iterative classifier outperforms N-over-D. As interval $\ell$ increases from $1$ to $10$ min, it significantly improves the iterative classifier's detection ability. There is a notable increase in performance from $\ell=1$ to $\ell=2$ min. However, increasing intervals $\ell$ did not affect N-over-D much, and the performance is not sensitive to this online-learning parameter.

\begin{table}[htbp]
  \centering
  \caption{Online performance comparison for values of interval $\ell$ on the \textit{CICIDS2017} dataset---LOIC attack.}
  \resizebox{\linewidth}{!}{%
  \begin{tabular}{c||c|ccccc}
    \hline
    \textbf{Model} &\textbf{$\ell$} (min) 
    &\textbf{ACC} &\textbf{FPR} & \textbf{Prec.} & \textbf{Rec.} & \textbf{F1} \\ \hline
    \multirow{4}{*}{\begin{tabular}[c]{@{}l@{}}Iterative \\ Classifier\end{tabular}} &$1$   &0.8805	&0.0329	&0.8808	&0.6494	&0.7476 \\   
    &$2$    &0.9355	&0.0340	&0.8529	&0.8104	&0.8311 \\
    &$5$    &0.9462	&0.0165	&0.9212	&0.7932	&0.8524 \\
    &$10$   &0.9486	&0.0220	&0.9016	&0.8275	&0.8630   \\ \hline    
    \multirow{4}{*}{\begin{tabular}[c]{@{}l@{}}N-over-D \end{tabular}} &$1$ &0.9330	&0.0047	&0.9959	&0.8846	&0.9370    \\
    &$2$ &0.9362	&0.0009	&0.9992	&0.8875	&0.9400 \\
    &$5$ &0.9371	&0.0001	&0.0999	&0.8882	&0.9408    \\
    &$10$ &0.9338	&0.0037	&0.9967	&0.8853	&0.9377    \\
  \end{tabular}
}
\label{tab:lparams}
\end{table}

\begin{figure*}[htbp]
    \centering
    \resizebox{.8\linewidth}{!}{%
        \begin{minipage}[b]{.49\textwidth}
        \centering
        \subfloat[][N and Offline D]{
            \includegraphics[width=\linewidth]{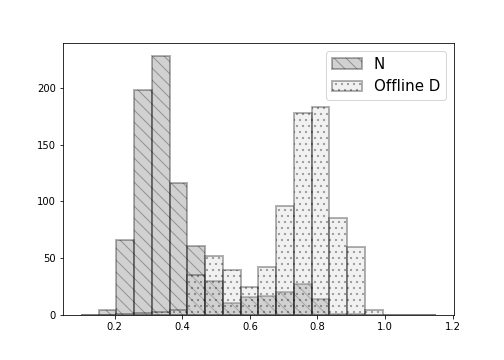}}
        \end{minipage} 
        \begin{minipage}[b]{.49\textwidth}
        \centering
            \subfloat[][N and Online D]{
            \includegraphics[width=\linewidth]{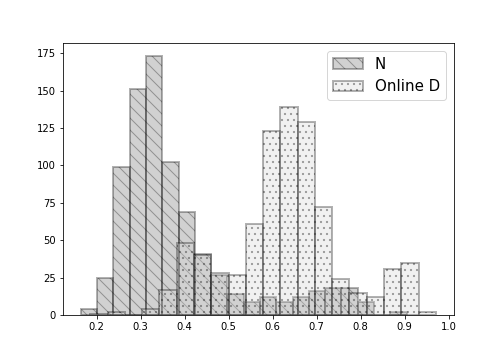}}
        \end{minipage}
    }
    \caption{The distinct distributions of the LOIC attack traffic determined by models N and D. This divergence enable N-over-D to identify the DDoS attacks. The x-axis is the normalized score values (0--1), and the y-axis is the number of traffic sequences.}
    \label{fig:ND_distr}
    \end{figure*}

Next, in Figure~\ref{fig:ND_distr}, we examine the differences in the normalized sequence score values of \textit{LOIC} attack traffic. There is a clear distinction between the distribution of the $N$ predicted values and the values expected by $D$ of the same traffic. It can be seen that $N$ assigns lower values for the attack traffic as it predicts the likelihood of the traffic as normal. On the other hand, both online and offline $D$ assigns higher scores to the same attack traffic as it predicts a high likelihood for these sequences to be attacks. Hence, by taking the $\frac{N}{D}$ scores, we obtain a more accurate measure of the probability that a sequence of traffic is normal, with a higher score indicating legitimacy.

Our N-over-D approach provides effective cover with practical considerations, but it is not without limitations. Due to the use of LSTM recurrent neural networks and sequences, it requires some time to collect the traffic data in order to form them into sequences before we begin the learning and mitigation process. This could potentially cause an undesired delay, which is easily circumvented by using another type of training algorithms instead of the LSTM in this framework. Also, since accurately estimating the attack and normal distribution difference is fundamental in retrieving the $\frac{N}{D}$ scores, our approach is most effective on volumetric DDoS attacks. It is not designed for low-rate attacks, and we have yet to study the performance of N-over-D on these types of attacks.

% ...


% ...

%% file: section/sec8-related.tex
\section{Related Work}
\label{sec:relatedwork}
In this section, we examine some of the most relevant works that take statistical and machine learning approaches in DDoS defenses.

\subsection{Statistical Approaches to DDoS Defense}
\label{subsec:statistical}
There is a large body of work on statistical approaches to DDoS mitigation since the early $2000$s~\cite{6489876}. Common methods generally involve some statistical measure of network traffic properties, such as entropy scoring of network packets~\cite{7345297,6601602,8256833,8466805}, IP filtering~\cite{1204223,Peng02detectingdistributed}, and IP source traceback~\cite{10.1007/3-540-36159-6_4,5467062}. These statistical and packet-based defense mechanisms are widely popular in defending against DDoS attacks. 
Statistical methods leverage variations of distributions in traffic attributes to discriminate between the distribution of traffic during normal usage and attack periods, and identify DDoS attacks.

Entropy-based detection approaches generally score packets based on some statistical metrics. In one of such works~\cite{6601602}, it proposes a semi-Markov model for normal network browsing behavior. Through simulating the browsing dynamics of legitimate web browsers, the authors observed that a critical condition for successful attacks: the number of active bots of the botnet must not be lower than the number of active legitimate users. Based on the findings, they defined a new fine correntropy metric to detect DDoS attacks. Another proposed method~\cite{8256833} combines an entropy-based and a packet-score-based method for first characterizing incoming packets before filtering the malicious packets. Results show that the combined scoring method can differentiate between DDoS attacks and normal traffic. Similarly, a statistical solution~\cite{8466805} proposes a joint entropy-based security scheme against DDoS attacks. It generates baseline profiles, determines parameters when attacks are detected, and mitigates the DDoS attacks.

IP filtering, based on IP source address filtering, is another practical way to defend against DDoS attacks. Unlike methods that are based on monitoring the traffic volume, the IP filtering scheme~\cite{1204223} uses a history of legitimate IP addresses to decide if incoming IP packets are anomalous. The authors present several heuristic methods to make the IP address database accurate and robust to increase the effectiveness of the scheme. Peng et al.~\cite{Peng02detectingdistributed} uses a sequential nonparametric change point detection method to monitor the increase of new IP addresses, and demonstrate its effectiveness for DDoS attacks. IP source traceback methods tackle the problem of identifying the DDoS attack sources. As monitoring the traffic attributes and IP filtering could provide limited information to prevent DDoS attacks, traceback methods~\cite{10.1007/3-540-36159-6_4,5467062} are another approach for effective defense against these attacks.

A common obstacle to these statistical techniques is the selection of an appropriate model parameters. Given that each DDoS attack in different networks varies from one to the next, it is especially challenging to identify the most suitable parameters that minimize false detection rates. Himura et al.~\cite{5198722} presented an automated parameter tuning of a statistical network traffic anomaly detection method, which determines a learning period for setting a parameter of the method. 

However, many of the statistical solutions may not be appropriate in an online setting, as some traffic statistics cannot be computed during data collection periods. To overcome this limitation, a DDoS defense architecture~\cite{1354679} was proposed to support distributed detection and automated online attack characterization, which assigns a score to each packet that estimates the legitimacy of a packet given the attributes.

\subsection{Machine Learning for DDoS Defense}
\label{subsec:ml}
Another popular approach to DDoS defense is based on Machine Learning (ML) anomaly detection techniques. Learning algorithms such as the Support Vector Machine (SVM), Self-Organizing Map (SOM), Random Forest, K-Nearest Neighbor (KNN), Decision Tree, Gaussian EM, and Neural Network have achieved various degrees of success in network intrusion detection~\cite{5670479,7987186}. Fundamental methods SOM~\cite{5735752} and SVM~\cite{7562606, 7816865, 8630919} were first proposed to provide artificial intelligence in DDoS mitigation. Braga et al.~\cite{5735752} presented a SOM detection scheme with 4 and 6 tuples of attributes as a lightweight detection method based on traffic flow features. Another work~\cite{7562606} applied the SVM classifier for detecting attacks. Some others~\cite{7816865,8630919} took a hybrid approach that combined the SVM and SOM to enhance the accuracy in differentiating normal flows from abnormal flows.

A proposed method~\cite{5670479} employs a one hidden layer feedforward neural network to detect abnormal traffic, and it achieves high accuracy and low false detection. In another work~\cite{7987186}, the authors evaluate nine ML methods on the source side data in the cloud for DDoS attack detection. The features (e.g., DNS packages ratio, Dillie-Hellman key exchange packages, ICMP package rate) are first extracted from the traffic data. The extracted data is then fed into the ML algorithms to distinguish between benign and malicious network traffic. Even though these ML methods are promising, the reported results depend heavily on the selection of features and the evaluated datasets~\cite{8441286}. Hence, to improve the generalization of traffic data and reduce the significance of feature engineering, deep learning models have been increasingly used in recent years.

The recent advancement of machine learning has given rise to increasing exploration of deep learning (DL) models for DDoS detection. We can broadly categorize these DL models into two groups---those that adopt an offline learning approach~\cite{8066291,10.1007/978-3-030-00018-9_15,doi:10.1002/dac.3497,8416441,Yuan2017DeepDefenseID,8666588}, where all attack data is observed and available for training the models, and online learning models~\cite{lima2019smart,7987186,DANESHGADEHCAKMAKCI2020102756,Doriguzzi_Corin_2020} that only receive data, as more attack traffic is observed with time.

One offline method~\cite{8416441}, CNN models with different model depths 
were used to investigate the relationship between performance and the number of CNN layers. They observe that deeper structures do not improve performance, and the LSTM model, a variant of RNN, performs better. The modeling of network traffic through time in a sequential manner, which retains information of historical patterns, instead of treating the inputs as independent packets. RNN-based intrusion detection systems~\cite{8066291,10.1007/978-3-030-00018-9_15,doi:10.1002/dac.3497} demonstrate that the sequential learning models have superior modeling capabilities, and they can detect some sophisticated attacks for intrusion detection tasks. 
RNN-based methods have demonstrated superior detection capabilities on some sophisticated attacks for intrusion detection tasks~\cite{8066291,10.1007/978-3-030-00018-9_15,doi:10.1002/dac.3497}.
In addition, a few deep learning methods~\cite{Yuan2017DeepDefenseID,8666588} that experiment with RNNs and CNNs hybrid models have slightly increased DDoS detection performance. However, all these methods are not designed for online systems, since the entire training data must be preprocessed before the models start training. They are not suitable for incremental updates when new data is observed.


While there are many offline learning-based DDoS detection methods, few online approaches have been proposed. One online method~\cite{Doriguzzi_Corin_2020} that uses ground-truth labels for training, extracts traffic flow attributes (e.g., Packet Length, TCP Length, Window Size) as input for a deep CNN model.

Only a few methods do not use data labels for training. He et al.~\cite{7987186} evaluated nine basic ML methods with an added online learning module on the source side data in the cloud for DDoS attack detection. In another proposed online approach~\cite{lima2019smart}, basic learning algorithms and simple decision tree methods were employed. Last but not least, {Çakmakçı} et al.~\cite{DANESHGADEHCAKMAKCI2020102756} employs a kernel-based learning algorithm in addition to the Mahalanobis distance and a chi-square test. It was shown to be competitive with offline DDoS classification algorithms. 
However, these existing online approaches are defined by their limited modeling capabilities, in which deep learning methods are capable of overcoming the limitations.


%% file: section/sec9-conclude.tex
\section{Conclusion}
\label{sec:conclude}
This work presents a principled formulation of a machine learning optimization problem that optimizes the detection of attacks in DDoS situations, and it proposes two online learning approaches.
First, we introduce N-over-D, an LSTM-based training algorithm that mitigates DDoS attacks by contrasting estimated normal and attack network traffic conditional probability distributions and ranking the unidentified traffic. However, this approach requires an exact likelihood calculation, and there is no joint training of the $N$ and $D$ models.
Second, we propose an enhanced iterative two-class classifier and design a specific loss function more suited for deep learning that solves the presented optimization problem.
Through extensive evaluation, we demonstrate the effectiveness of N-over-D and iterative classifier against a range of DDoS attacks, actual and synthetically generated, offering a practical solution to the ever-evolving DDoS attacks. We recognize the lack of ground truth labels that identify traffic during the attacks, which contains a mixture of legitimate and malicious traffic, raising serious concern about the effectiveness of existing machine learning detection methods that rely on these labels for the reported performance. 
Furthermore, we analyze the strengths and weaknesses of both approaches, which illustrate the practical considerations of designing a more functional and robust online DDoS mitigation system.

%% file: section/appendix.tex

\section{Implementation Details} 
We provide the LSTM architecture operations, describe the network traffic data preprocessing procedure, and report the default parameter settings for the experiments implemented in this paper.

\subsection{LSTM Architecture Computation}

We use the LSTM architecture without peep-hole connections. The operations are computed as follows:

\begin{align*}
  &\begin{aligned}
    i_t &= \sigma(W_{xi}x_t + W_{hi}h_{t-1} + b_i)  
  \end{aligned}\\
  &\begin{aligned}
    f_t &= \sigma(W_{xf}x_t + W_{hf}h_{t-1} + b_f)
  \end{aligned}\\
  &\begin{aligned}
    o_t &= \sigma(W_{xo}x_t + W_{ho}h_{t-1} + b_o)  
  \end{aligned}\\
  &\begin{aligned}
    g_t &= \tanh(W_{xg}x_t + W_{hg}h_{t-1} + b_g) 
  \end{aligned}\\
  &\begin{aligned}
    c_t &= f_t \odot c_{t-1} + i_t \odot g_t  
  \end{aligned}\\
  &\begin{aligned}
    h_t &= o_t \odot \tanh(c_t) 
  \end{aligned}
\end{align*}
where $x_t$ is the input at time $t$, $\sigma(\cdot)$ is the sigmoid function, $\tanh(\cdot)$ is the hyperbolic tangent function, and $\odot$ denotes element-wise product. Each gate function has its own weight matrix and a bias vector. We denote the parameters with subscripts $f$ for the forget gate function, $i$ for the input gate function, and $o$ for the output gate function respectively (e.g., $W_{xf}$, $W_{hf}$, and $b_f$ are parameters of the forget gate function).

\subsection{Data Preprocessing Algorithm}

Using the data processing procedure below, we prepare network traffic data into a streaming form, grouped by source and destination IP addresses, for online learning.

\begin{figure}[htbp]
\centering
\resizebox{.99\linewidth}{!}{ 
\begin{minipage}{\linewidth}
\begin{algorithm}[H]
\caption{Network data preprocessing algorithm.
}\label{alg:preprocess}
\begin{algorithmic}[1]
    \For{each interval $\tau_i$} 
    \State Sort packets $\bm{x} = \{x_{1}, \dots, x_{n}\}$ chronologically by absolute time
    \ForAll{$x \in  \bm{x}$}
    \If{number of packets $x^{(ip)}$ from an IP  $< \varepsilon$} 
    \State $\bm{x} \setminus x^{(ip)}$
    \EndIf
    \State Extract features $\bm{f}$
    \EndFor
    \State Form sequences $seq^{(ip)} = (f^d_1, f^d_2, \dots, f^d_T)$ with dynamic features of length $seq\_len = T$ 
    \If{$seq\_len < T$} 
    \State $zero\_pad(seq)$
    \EndIf
    \State $seq^{(ip)} \cup f^{\bm{s}(ip)}$  \algorithmiccomment{Join static features of IP to sequences}
    \EndFor   %
\end{algorithmic}
\end{algorithm}  
\end{minipage}
}
\end{figure}

\subsection{Parameter Settings} 
For the numerator model $N$, we split the training data into an 80/20 training and validation ratio. As for the online model $D$, every batch of data in a time interval is divided into a 90/20 training and validation ratio to maximize the training data. \cref{tab:NDparams} lists the parameter settings of numerator $N$.

\begin{table}[htbp]
  \centering
  \caption{Model Parameters.}
  \resizebox{\linewidth}{!}{%
  \begin{tabular}{c||c|c|c}
    Dataset & Parameter & $N$ & $D$ \\ \hline 
    \multirow{5}{*}{CAIDA07}    &epochs   &30   &10 (1 per interval)   \\ \cline{2-4}
    &learning rate    &0.005   &0.003   \\ \cline{2-4}
    &LSTM dimension    &300   &300   \\ \cline{2-4}
    &embedding dimension    &512   &512   \\ \cline{2-4}
    &batch size    &512   &128   \\ \cline{2-4}
    &sequence length    &200  &200   \\ \hline
    \multirow{5}{*}{LOIC}    &epochs   &30   &10 (5 per interval)  \\ \cline{2-4}
    &learning rate    &0.005   &0.003   \\ \cline{2-4}
    &LSTM dimension    &300   &300   \\ \cline{2-4}
    &embedding dimension    &512   &512   \\ \cline{2-4}
    &batch size    &512   &128   \\ \cline{2-4}
    &sequence length    &200  &200   \\ \hline    
    \multirow{5}{*}{HULK}    &epochs   &30   &10 (5 per interval)  \\ \cline{2-4}
    &learning rate    &0.005   &0.003   \\ \cline{2-4}
    &LSTM dimension    &300   &300   \\ \cline{2-4}
    &embedding dimension    &512   &512   \\ \cline{2-4}
    &batch size    &512   &128   \\ \cline{2-4}
    &sequence length    &90  &90   \\ \hline       
  \end{tabular}
}
\label{tab:NDparams}
\end{table}

\section{Prediction Analysis}
\begin{figure}[htbp]
        \centering
        \includegraphics[width=.9\linewidth]{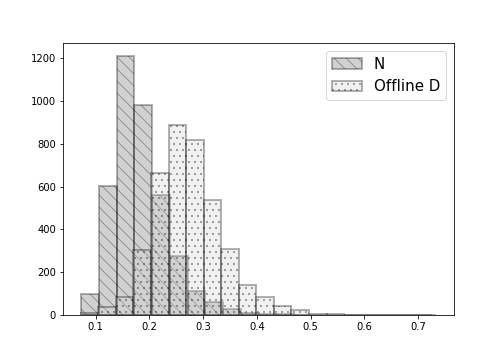}
        \caption{HULK attack. Prediction by $N$ and offline $D$.}
        \label{fig:hlk_offline}
    \end{figure}

\begin{figure}[htbp]
        \centering
        \includegraphics[width=.9\linewidth]{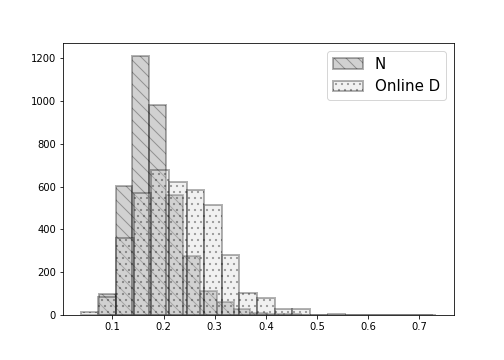}
        \caption{HULK attack. Prediction by $N$ and online $D$.}
        \label{fig:hlk_online}
    \end{figure}

\begin{figure}[htbp]
        \centering
        \includegraphics[width=.9\linewidth]{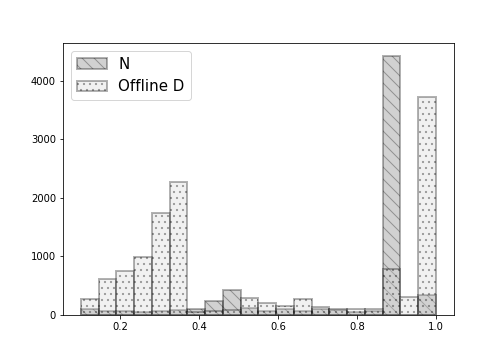}
        \caption{CAIDA07 attack. Prediction by $N$ and offline $D$.}
        \label{fig:caida_offline}
    \end{figure}

\begin{figure}[htbp]
        \centering
        \includegraphics[width=.9\linewidth]{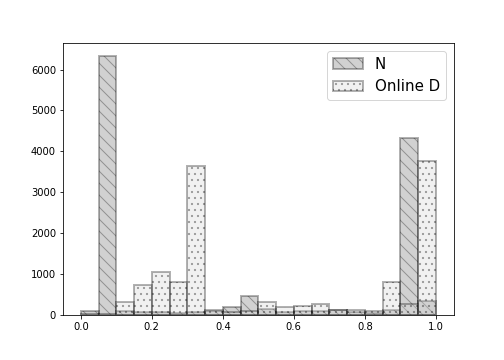}
        \caption{CAIDA07 attack. Prediction by $N$ and online $D$.}
        \label{fig:caida_online}
    \end{figure}

We include more experimental analysis below that illustrate the different distributions learned by models $N$ and both the offline and online $D$ models. \cref{fig:hlk_offline,fig:hlk_online} show the divergent prediction distributions for the HULK attack. For the distributions of the predictions for the CAIDA07 attack, refer to \cref{fig:caida_offline,fig:caida_online}.


